\documentclass[aps,tightenlines,superscriptaddress,twocolumn]{revtex4-1}
\usepackage{graphicx}
\usepackage{bm}
\usepackage{times}
\usepackage{lineno}
\usepackage[pscoord]{eso-pic}
\usepackage{amsmath}

\newcommand{\snn}{$\sqrt{s_{\rm NN}}$ }

\begin{document}

\title{
Disappearance of partonic collectivity in \snn = 3\,GeV Au+Au collisions at RHIC}

\affiliation{Abilene Christian University, Abilene, Texas   79699}
\affiliation{AGH University of Science and Technology, FPACS, Cracow 30-059, Poland}
\affiliation{Alikhanov Institute for Theoretical and Experimental Physics NRC "Kurchatov Institute", Moscow 117218, Russia}
\affiliation{Argonne National Laboratory, Argonne, Illinois 60439}
\affiliation{American University of Cairo, New Cairo 11835, New Cairo, Egypt}
\affiliation{Brookhaven National Laboratory, Upton, New York 11973}
\affiliation{University of California, Berkeley, California 94720}
\affiliation{University of California, Davis, California 95616}
\affiliation{University of California, Los Angeles, California 90095}
\affiliation{University of California, Riverside, California 92521}
\affiliation{Central China Normal University, Wuhan, Hubei 430079 }
\affiliation{University of Illinois at Chicago, Chicago, Illinois 60607}
\affiliation{Creighton University, Omaha, Nebraska 68178}
\affiliation{Czech Technical University in Prague, FNSPE, Prague 115 19, Czech Republic}
\affiliation{Technische Universit\"at Darmstadt, Darmstadt 64289, Germany}
\affiliation{ELTE E\"otv\"os Lor\'and University, Budapest, Hungary H-1117}
\affiliation{Frankfurt Institute for Advanced Studies FIAS, Frankfurt 60438, Germany}
\affiliation{Fudan University, Shanghai, 200433 }
\affiliation{University of Heidelberg, Heidelberg 69120, Germany }
\affiliation{University of Houston, Houston, Texas 77204}
\affiliation{Huzhou University, Huzhou, Zhejiang  313000}
\affiliation{Indian Institute of Science Education and Research (IISER), Berhampur 760010 , India}
\affiliation{Indian Institute of Science Education and Research (IISER) Tirupati, Tirupati 517507, India}
\affiliation{Indian Institute Technology, Patna, Bihar 801106, India}
\affiliation{Indiana University, Bloomington, Indiana 47408}
\affiliation{Institute of Modern Physics, Chinese Academy of Sciences, Lanzhou, Gansu 730000 }
\affiliation{University of Jammu, Jammu 180001, India}
\affiliation{Joint Institute for Nuclear Research, Dubna 141 980, Russia}
\affiliation{Kent State University, Kent, Ohio 44242}
\affiliation{University of Kentucky, Lexington, Kentucky 40506-0055}
\affiliation{Lawrence Berkeley National Laboratory, Berkeley, California 94720}
\affiliation{Lehigh University, Bethlehem, Pennsylvania 18015}
\affiliation{Max-Planck-Institut f\"ur Physik, Munich 80805, Germany}
\affiliation{Michigan State University, East Lansing, Michigan 48824}
\affiliation{National Research Nuclear University MEPhI, Moscow 115409, Russia}
\affiliation{National Institute of Science Education and Research, HBNI, Jatni 752050, India}
\affiliation{National Cheng Kung University, Tainan 70101 }
\affiliation{Nuclear Physics Institute of the CAS, Rez 250 68, Czech Republic}
\affiliation{Ohio State University, Columbus, Ohio 43210}
\affiliation{Institute of Nuclear Physics PAN, Cracow 31-342, Poland}
\affiliation{Panjab University, Chandigarh 160014, India}
\affiliation{Pennsylvania State University, University Park, Pennsylvania 16802}
\affiliation{NRC "Kurchatov Institute", Institute of High Energy Physics, Protvino 142281, Russia}
\affiliation{Purdue University, West Lafayette, Indiana 47907}
\affiliation{Rice University, Houston, Texas 77251}
\affiliation{Rutgers University, Piscataway, New Jersey 08854}
\affiliation{Universidade de S\~ao Paulo, S\~ao Paulo, Brazil 05314-970}
\affiliation{University of Science and Technology of China, Hefei, Anhui 230026}
\affiliation{Shandong University, Qingdao, Shandong 266237}
\affiliation{Shanghai Institute of Applied Physics, Chinese Academy of Sciences, Shanghai 201800}
\affiliation{Southern Connecticut State University, New Haven, Connecticut 06515}
\affiliation{State University of New York, Stony Brook, New York 11794}
\affiliation{Instituto de Alta Investigaci\'on, Universidad de Tarapac\'a, Arica 1000000, Chile}
\affiliation{Temple University, Philadelphia, Pennsylvania 19122}
\affiliation{Texas A\&M University, College Station, Texas 77843}
\affiliation{University of Texas, Austin, Texas 78712}
\affiliation{Tsinghua University, Beijing 100084}
\affiliation{University of Tsukuba, Tsukuba, Ibaraki 305-8571, Japan}
\affiliation{Valparaiso University, Valparaiso, Indiana 46383}
\affiliation{Variable Energy Cyclotron Centre, Kolkata 700064, India}
\affiliation{Warsaw University of Technology, Warsaw 00-661, Poland}
\affiliation{Wayne State University, Detroit, Michigan 48201}
\affiliation{Yale University, New Haven, Connecticut 06520}

\author{M.~S.~Abdallah}\affiliation{American University of Cairo, New Cairo 11835, New Cairo, Egypt}
\author{B.~E.~Aboona}\affiliation{Texas A\&M University, College Station, Texas 77843}
\author{J.~Adam}\affiliation{Brookhaven National Laboratory, Upton, New York 11973}
\author{L.~Adamczyk}\affiliation{AGH University of Science and Technology, FPACS, Cracow 30-059, Poland}
\author{J.~R.~Adams}\affiliation{Ohio State University, Columbus, Ohio 43210}
\author{J.~K.~Adkins}\affiliation{University of Kentucky, Lexington, Kentucky 40506-0055}
\author{G.~Agakishiev}\affiliation{Joint Institute for Nuclear Research, Dubna 141 980, Russia}
\author{I.~Aggarwal}\affiliation{Panjab University, Chandigarh 160014, India}
\author{M.~M.~Aggarwal}\affiliation{Panjab University, Chandigarh 160014, India}
\author{Z.~Ahammed}\affiliation{Variable Energy Cyclotron Centre, Kolkata 700064, India}
% ...

\collaboration{STAR Collaboration}\noaffiliation

%\linenumbers

\begin{abstract}
 
 We report on the measurements of directed flow $v_1$ and elliptic flow $v_2$ for hadrons ($\pi^{\pm}$, $K^{\pm}$, $K_{S}^0$, $p$, $\phi$, $\Lambda$ and $\Xi^{-}$) from Au+Au collisions at $\sqrt{s_{NN}}$ = 3\,GeV and $v_{2}$ for ($\pi^{\pm}$, $K^{\pm}$, $p$ and $\overline{p}$) at 27 and 54.4\,GeV with the STAR experiment. While at the two higher energy midcentral collisions the number-of-constituent-quark (NCQ) scaling holds, at 3\,GeV the $v_{2}$ at midrapidity is negative for all hadrons and the NCQ scaling is absent. In addition, the $v_1$ slopes at midrapidity for almost all observed hadrons are found to be positive, implying dominant repulsive baryonic interactions. The features of negative $v_2$ and positive $v_1$ slope at 3\,GeV can be reproduced with a baryonic mean-field in transport model calculations. These results imply that the medium in such collisions is likely characterized by baryonic interactions.

\end{abstract}

\maketitle

\hyphenpenalty=700
\tolerance=100
%%%%%%%%%%%%%%%%%%%%%%%%%%%%%%%%%%%%%%%%%%%%%%%%%%%%%%%%%%%%%%%%%%%%%%%%%%%%%%%%%%%%%%%%%%%%%%%%%%%%%%%%%
\section{Introduction}
Relativistic heavy-ion collisions at the Large Hadron Collider (LHC) and the Relativistic Heavy Ion Collider (RHIC), where the net-baryon density is low, are generally considered to have produced a new form of matter with partonic degrees of freedom, usually referred to as the strongly-coupled Quark Gluon Plasma (sQGP)~\cite{Arsene:2004fa,Adcox:2004mh,Back:2004je,Adams:2005dq,Aamodt:2010pa}. 
However, it is necessary to identify changes in physical properties, \textit{e.g.} in its equation of state (EOS), before ultimately claiming the discovery of the new form of matter.
Since the discovery of the sQGP in 2005, the nature of the phase transition from hadronic matter to the QGP and of the Quantum Chromodynamics (QCD) phase diagram at finite net-baryon density have been the focus in the RHIC beam energy scan program.
This is, after the discovery of the sQGP at vanishing net-baryon density, an important step toward understanding the phase structure of nuclear matter in the high baryon density region.

In order to extract underlying dynamic information, the particle differential distribution is often written in the form of a Fourier series~\cite{PhysRevLett.78.2309,Sorge:1997nv,Ollitrault:1992bk},
\begin{equation}
E\frac{d^3N}{d^3p}=\frac{1}{2\pi}\frac{d^2N}{p_Tdp_Tdy}(1+\sum_{n=1}^{\infty}2v_n{\rm{cos}}(n(\phi-\Psi)))
\end{equation}
where $p_T$, $y$, $\phi$ and $\Psi$ are, respectively, particle transverse momentum, rapidity, azimuthal angle of the particle and the event plane angle. 
Due to their sensitivity to the expansion dynamics of the produced matter, the first two Fourier expansion coefficients $v_{1}$ (directed flow) and $v_{2}$ (elliptic flow) are sensitive probes for studying the properties of the matter created in high-energy nuclear collisions~\cite{Hung:1994eq,Steinheimer:2014pfa,Nara:2016phs}.
At higher energies (nucleon-nucleon center-of-mass energy \snn $\gtrsim$ 27\,GeV), where the transit time of the colliding nuclei ($ \sim 2R/\gamma\beta$) is smaller than the typical production time of  particles~\cite{Bialas:1986cf,PhysRevC.98.034908}, flow harmonics are dominated by the collective expansion of initial partonic density distribution~\cite{Abelev:2007rw,Adamczyk:2015ukd,Adamczyk:2017xur}. At lower energies, shadowing effect by the passing spectator nucleons becomes
important~\cite{Liu:2000am,Pinkenburg:1999ya,Adamczyk:2017nxg,Adamczyk:2014ipa,Adamczyk:2013gw, Adam:2020pla}. At \snn $\lesssim$ 4\,GeV, nuclear mean-field effects will contribute to the observed azimuthal anisotropies~\cite{Buss:2011mx,Bass:1998ca,Bleicher:1999xi,Andronic:2004cp}. Previous studies have shown that $v_1$ and $v_2$ are particularly sensitive to the incompressibility ($\kappa$) of the nuclear matter in the high baryon density region~\cite{Danielewicz:2002pu,Danielewicz:1998vz,Kruse:1985hy,Wang:2018hsw}. The constraints on $\kappa$ by comparing experimental data with results from the theoretical transport model will certainly help us to understand nuclear EOS. In a systematic analysis of hadron spectra and anisotropic flow in Au+Au collisions at the energy range of \snn = 2 – 4.5 GeV, the authors concluded that anisotropic flow is sensitive to the EOS and a realistic EOS with a transition to QGP is needed in order to understand the experimental observations in the high baryon density region~\cite{Spieles:2020zaa}.

Large positive $v_2$, especially for multistrange hadrons,  along with the observation of its number-of-constituent-quarks (NCQ) scaling are strong evidence for the formation of a hydrodynamically expanding QGP phase with partonic degrees of freedom~\cite{Abelev:2007rw,Adamczyk:2015ukd,Adamczyk:2017xur}. 
Positive $v_{2}$ of light hadrons at midrapidity has been observed from the top RHIC energy down to 4.5\,GeV~\cite{Adam:2020pla}.
On the other hand, at \snn $\ge$ 10\,GeV, all midrapidity $v_1$ slopes are found to have negative values and approach to zero with increasing energy~\cite{Adamczyk:2017nxg,Adamczyk:2014ipa,Brachmann:1999xt}, where partonic collectivity is dominant. At lower collision energies the $v_{1}$ slope values for baryons become positive, while those for mesons remain negative~\cite{Adam:2020pla,Kintner:1997ss,Liu:2000am,Chung:2000ny}. For example, results of proton and light nuclei $v_1$ and $v_2$ from Au+Au collisions at \snn = 2.4 GeV were reported recently by the HADES experiment~\cite{Adamczewski-Musch:2020iio}.

\section{Experiment and data analysis}
In this paper we report systematic results of $v_1$ and $v_2$ for identified hadrons ($\pi^{\pm}$, $K^{\pm}$, $K_{S}^0$, $p$, $\phi$, $\Lambda$, and $\Xi^{-}$) from 10-40\% centrality Au+Au collisions at \snn = 3\,GeV and $v_{2}$ of ($\pi^{\pm}$, $K^{\pm}$, $p$, and $\bar{p}$) at \snn $=$ 27 and 54.4\,GeV
from the STAR experiment. The data sets at 3, 27, and 54.4\,GeV are 260, 560, and 600 $\times$ $10^{6}$ events with minimum-bias trigger, respectively.
The main detector of STAR is a cylindrical Time Projection Chamber (TPC)~\cite{Anderson:2003ur} 4\,m in diameter and 4\,m in length. The TPC resides in a solenoidal magnet providing a uniform magnetic field of 0.5 T along the longitudinal beam direction. The data at 3 GeV were taken, with beam energy of 3.85\,GeV per nucleon, in 2018 in the fixed-target (FXT) mode. The target, with a thickness of 0.25\,mm corresponding to a 1\% interaction probability, is positioned inside the beam pipe near the edge of the TPC, at 200.7 cm from the TPC center along the beam axis. This gives an experimental acceptance coverage of $-2$ $< \eta <$ 0 in pseudorapidity in the lab frame. The higher energy data were taken in the collider mode, where the beam bunch crossing was restricted to the TPC central region, yielding an acceptance of $|\eta|<$ 1.

%%%%%%%%%%%%%% Fig. 1 %%%%%%%%%%%%%%%%%%%%
\begin{figure}[!htb]
\centering
\centerline{\includegraphics[width=0.5\textwidth]{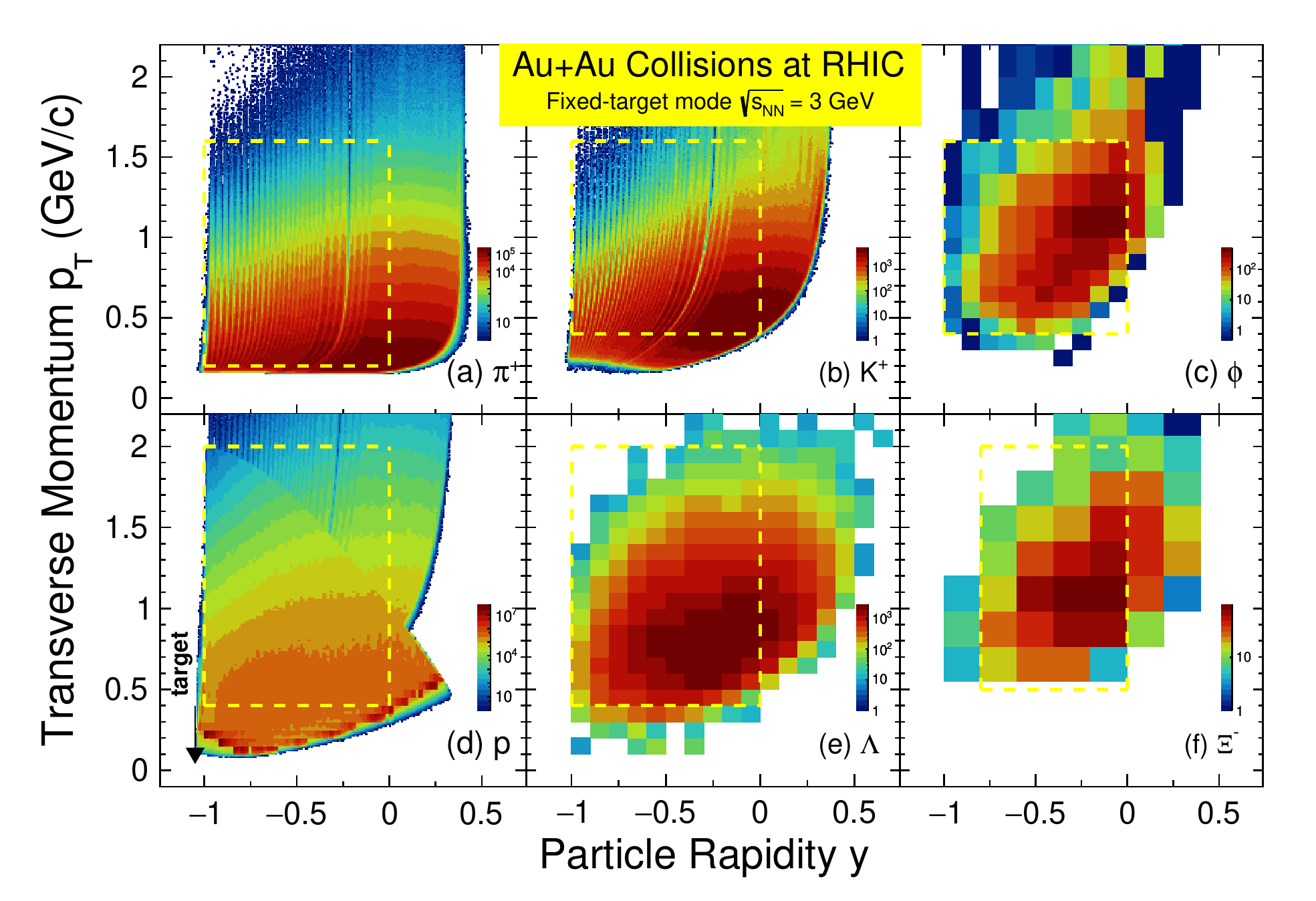}}
\caption{The efficiency uncorrected density distributions in transverse momentum ($p_{T}$) and identified particle rapidity ($y$) for $\pi^{+}$, $K^{+}$, $\phi$, $p$, $\Lambda$ and $\Xi^{-}$ measured with the STAR detector TPC and TOF in Au+Au collisions at \snn = 3\,GeV, with the FXT mode of beam energy 3.85\,GeV per nucleon. The target is located at $y$ = $-1.05$. In each plot, intensity is self-normalized.}
\label{fig1}
\end{figure}
%%%%%%%%%%%%%%%%%%%%%%%%%%%%%%%%%%%%%

%\textbf{\textit{Centrality Definition and Event, Track Cuts:}}
The centrality of collisions is characterized by the number of charged tracks detected with the TPC within pseudorapidity $|\eta|<$ 0.5 in collider mode collisions and $-2$ $<\eta<$ 0 for FXT mode collisions. When two (or more) independent single collision events are superposed, it is called pile-up which often occurs in the fixed target mode. In order to remove the pile-up effect, events with multiplicity greater than 195 are excluded from the analysis at \snn = 3\,GeV. 
The primary vertex position of each event along the beam direction, $V_{z}$, is required to be within $\pm$40 cm of the center of the TPC at \snn =  27 and 54.4\,GeV, and within $\pm$2 cm of the target position for the FXT mode collisions at \snn = 3\,GeV. An additional selection on the primary vertex position within a radius less than 2 cm is required to eliminate possible beam interactions with the vacuum pipe of 4 cm radius at all three energies. In order to improve the track quality, momentum and ionization energy loss resolution from the TPC, the following track selections are applied: i) the number of hit points is larger than 15; ii) the ratio between the number of hit points and the maximum possible number of hit points is larger than 0.52; iii) the distance of closest approach (DCA) to the primary vertex is less than 3 cm~\cite{Adamczyk:2013gw}. 

The particle identification of charged pions with transverse momentum range 0.2 $<p_{T}<$ 1.6\,GeV/$c$, charged kaons with 0.4 $<p_{T}<$ 1.6\,GeV/$c$, and protons with 0.4 $<p_{T}<$ 2.0\,GeV/$c$ are based on ionization energy loss information measured with the TPC detector and time-of-flight information measured with the Time-of-Flight (TOF) detector~\cite{Geurts:2004fn}. Reconstruction of $K_S^{0}$, $\Lambda$, and $\Xi^{-}$ is performed using the KF Particle Finder package based on the Kalman Filter method, initially developed for the CBM and ALICE experiments~\cite{Kisel:2018nvd}, and also used in STAR measurements~\cite{STAR:2020xbm}. In order to enhance the signal significance, the method utilizes the covariances of track parameters to determine and select on variables characterizing decay topology. The $\phi$ mesons are reconstructed through the decay channel, $\phi \rightarrow K^{+}+K^{-}$, where the combinatorial background is estimated using the mixed-event technique~\cite{Adamczyk:2013gw}.

Figure~\ref{fig1} presents the density distributions in $y$ and $p_{T}$ for $\pi^{+}$, $K^{+}$, $p$, $\phi$, $\Lambda$, and $\Xi^{-}$, measured with the TPC and TOF detectors in Au+Au collisions at \snn = 3\,GeV. In the remainder of this paper, all notations are presented in the center-of-mass frame for both the collider and FXT datasets. The target is located at $y$ = $-1.05$ and the positive sign of $v_1$ is defined by the forward positive rapidity region. The acceptance for all particles covers from midrapidity to target rapidity. The coverage of $p_{T}$ is from 0.2 to $\sim$ 2\,GeV/$c$, depending on the rest mass of the particle.

\begin{figure}[!htb]
\centering
\centerline{\includegraphics[width=0.5\textwidth]{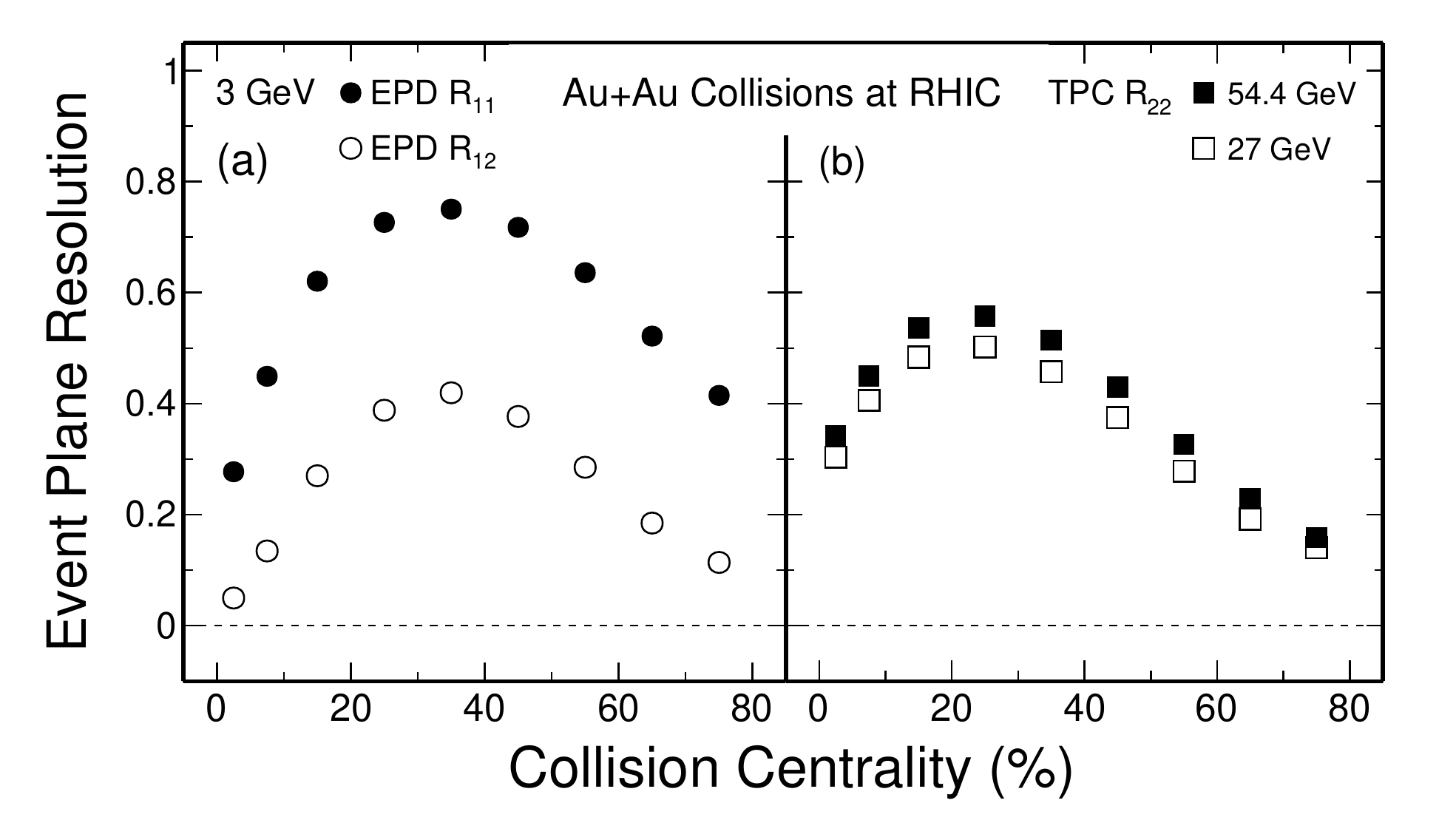}}
\vspace{-0.3cm}
\caption{Event plane resolution as a function of collision centrality from Au+Au collisions at \snn $=$ 3 (a), 27 and 54.4\,GeV (b). In case of the 3 GeV collisions, $\Psi_1$ is used to determine the event plane resolutions for the first and second harmonic coefficients shown as ${\rm R}_{11}$ and ${\rm R}_{12}$ in left panel. In the 27 and 54.4 GeV collisions, $\Psi_2$ is used to evaluate the second order event plane resolution, see right panel. In all cases, the statistic uncertainties are smaller than symbol sizes.}
\label{fig:epr}
\end{figure}

Before extracting the flow information one must determine the event plane angle on an event-by-event basis~\cite{STAR:2000ekf,Poskanzer:1998yz}. For the $n^{\rm th}$ Fourier harmonic, the flow vector $\vec{Q}_{n}$ = ($Q_{nx}, Q_{ny}$) and the event plane angle $\Psi_{n}$ are event-by-event calculated by
\begin{equation}
Q_{nx} = \sum_{i}w_{i}{\rm cos}(n\phi_{i}), Q_{ny} = \sum_{i}w_{i}{\rm sin}(n\phi_{i}), $$ $$\Psi_{n} = \left( {\rm tan^{-1}}\frac{Q_{ny}}{Q_{nx}} \right) / n
\end{equation}
where sums go over all particles $i$ used in the event plane calculation, $\phi_{i}$ is the laboratory azimuthal angle, and the weight $w_{i}$ used here is $p_{T}$ for the $i^{\rm th}$ particle.
%, and $\phi_{i}$ and $w_{i}$ are the laboratory azimuthal angle and the weight for the $i^{\rm th}$ particle, respectively.
For the Au+Au collisions at \snn $=$ 27 and 54.4\,GeV, the second order event plane angle ($\Psi_2$) is reconstructed with tracks determined by the TPC and the event plane resolution is determined as $\rm{R_{22}} = \sqrt{\langle {\rm{cos}2(\Psi^A_2-\Psi^B_2) \rangle}}$, where A and B are independent subevents, from $\eta$ ranges $-1$ $< \eta <$ $-0.05$ and $0.05$ $< \eta <$ $1$, respectively. The average $\langle ..\rangle$ runs over all events. 
%To avoid self-correlation and suppress non-flow effects, the $\eta$-subevent plane method~\cite{PhysRevC.58.1671} is used for the elliptic flow calculation, in which the $\eta$ ranges $-1$ $< \eta <$ $-0.05$ and $0.05$ $< \eta <$ $1$ are applied for the two independent subevents, separately. 
At \snn $=$ 3\,GeV, the first order event plane angle ($\Psi_1$) is determined with the Event Plane Detector (EPD) covering the pseudorapidity region of $-5.3<\eta<-2.6$~\cite{Adams:2019fpo} in the lab frame. In this case, due to the strong $v_{1}$ signal and better resolution from $\Psi_{1}$, a three-subevent method with both TPC and EPD is used to determine the first order event plane resolution ${\rm R_{11}} = \sqrt{\langle {\rm{cos}(\Psi^{A}_{1}-\Psi^{B}_{1})}\langle {\rm{cos}(\Psi^{A}_{1}-\Psi^{C}_{1})} /\langle{\rm{cos}(\Psi^{B}_{1}-\Psi^{C}_{1})} }$ 
for the $v_{1}$ measurements. The ${\rm R_{12}}$ is converted ${\rm R_{11}}$ for the measurements of $v_{2}$~\cite{Poskanzer:1998yz}. The resulting event plane resolution as a function of the collision centrality, is shown in Fig.~\ref{fig:epr}.
Using $\Psi_{1}$ to analyze $v_{2}$ is also used in the recent publication from HADES collaboration~\cite{Adamczewski-Musch:2020iio}.
In all cases, standard acceptance corrections are carried out to ensure a uniform distribution of the event plane angle~\cite{Poskanzer:1998yz}. The final results are corrected for centrality bin width, event plane resolution, tracking efficiency and detector acceptance~\cite{Adamczyk:2013gw,Adamczyk:2017nxg}.

%%%%%%%%%%%%%% Fig. 2 %%%%%%%%%%%%%%%%%%%%
\begin{figure*}[!htb]
\centering
\centerline{\includegraphics[width=0.8\textwidth]{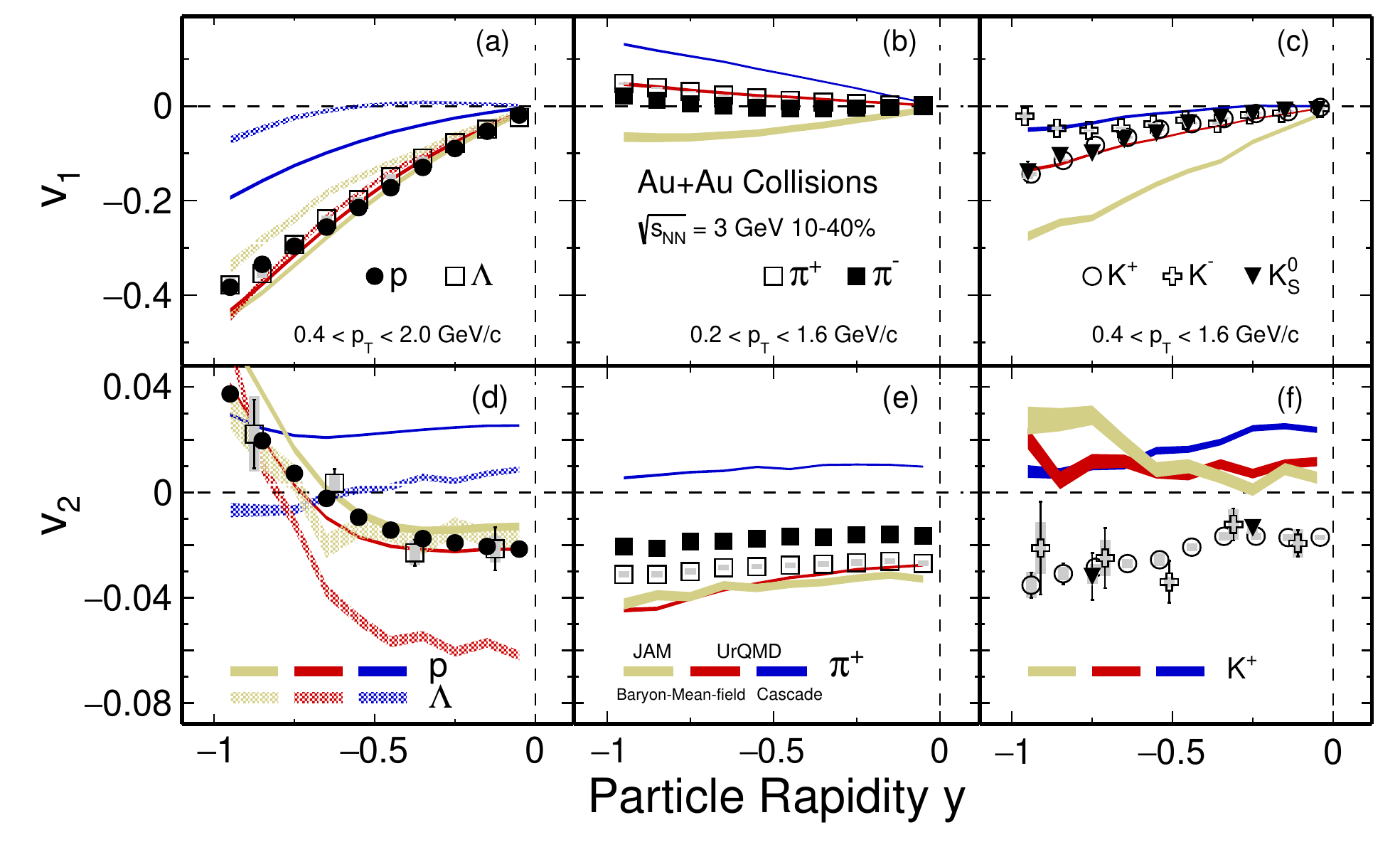}}
\caption{Rapidity($y$) dependence of $v_{1}$ (top panels) and $v_{2}$ (bottom panels) of proton and $\Lambda$ baryons (left panels), pions (middle panels) and kaons (right panels) in 10-40\% centrality for Au+Au collisions at \snn = 3\,GeV. Statistical and systematic uncertainties are shown as bars and gray bands, respectively. Some uncertainties are smaller than the data points. The UrQMD and JAM results are shown as bands: golden, red and blue bands stand for JAM mean-field, UrQMD mean-field and UrQMD cascade mode, respectively. The value of the incompressibility $\kappa$ = 380 MeV is used in the mean-field option. 
%More detailed model descriptions and data comparisons can be found in Supplemental Material.
}
\label{fig2}
\end{figure*}
%%%%%%%%%%%%%%%%%%%%%%%%%%%%%%%%%%%%%

Systematic uncertainties are estimated point-by-point by varying track selection criteria, and the decay length of parent and daughter when using the KF Particle Finder package~\cite{Kisel:2018nvd}.
At \snn = 3\,GeV, the leading systematic source is from particle misidentification by varying the ionization energy loss $dE/dx$, estimated to contribute 4.3\% (1.5\%) to $\pi^{+}$ (proton) $v_1$ slopes measurements. 
An additional, common systematic uncertainty from event plane resolution is estimated to be 1.4\% and 3\% for $v_1$ and $v_{2}$, respectively. Assuming the sources are uncorrelated, the total systematic uncertainty is obtained by adding uncertainties mentioned above in quadrature.

\section{Results and Discussions}
The rapidity dependence of the directed flow $v_1$ and elliptic flow $v_2$ of identified hadrons from Au+Au collisions at \snn = 3\,GeV in 10-40\% centrality is presented in Fig.~\ref{fig2}. Due to the acceptance, the results from the rapidity region $-1$ $\textless$ $y$ $\textless$ 0 are shown. 
The corresponding $p_T$ range for each hadron is shown in the figure. For comparison, calculations of transport theoretical model, JET AA Microscopic Transportation Model (JAM)~\cite{PhysRevC.61.024901} and Ultra-relativistic Quantum Molecular Dynamics (UrQMD)~\cite{Bass:1998ca,Bleicher:1999xi}, are also given for the abundantly produced hadrons $\pi^{+}$, $K^{+}$, $p$, and $\Lambda$. The results from the cascade and baryonic mean-field modes of the JAM and UrQMD model are shown as colored bands. The same collision centrality and kinematic selection criteria as in the data are applied in the model calculations. 

 The values of the midrapidity slope, defined as $dv_{1}/dy|_{y=0}$, are the largest for protons and $\Lambda$s, see panel (a), and are close to zero for pions in panel (b). In panel (c), $dv_{1}/dy|_{y=0}$ are positive and have small charge dependence among kaons. The JAM and UrQMD mean-field calculation includes a Skyrme potential energy density function~\cite{Kruse:1985hy}. Comparing to the cascade mode, the repulsive interactions among baryons are enhanced via an additional mean-field option, resulting in a good agreement with experimental data. A similar conclusion can be drawn for the elliptic flow $v_2$. As shown in the lower panels of Fig.~\ref{fig2}, all of the measured midrapidity hadrons, ($|y| \le 0.5$) show negative values of $v_2$ implying an out-of-plane expansion in the collisions at 3\,GeV, contrary to the in-plane expansion in high energy collisions~\cite{Adamczyk:2015ukd,Adamczyk:2017xur}. Again, with the mean-field option with $\kappa=380$ MeV, the JAM and UrQMD model calculations qualitatively reproduce the rapidity dependence of $v_2$ for baryons and pions. Nevertheless, we note that the UrQMD model overpredicts the strength of $v_2$ for strange baryon $\Lambda$ and both JAM and UrQMD model fails to reproduce kaon $v_2$, see Fig.~\ref{fig2}. It is worth noting that, due to the strong influence of the Coulomb potential, the integrated $v_2$ of $\pi^{-}$ are all smaller than that of $\pi^{+}$ over the measured rapidity range. In the above transport model calculations, no Coulomb force is included.

Similar to the previous $v_{1}$ studies~\cite{Adamczyk:2011aa,Adamczyk:2014ipa,Adamczyk:2017nxg} from the STAR experiment, a polynomial fit of the form $v_{1}(y) = a + by+cy^{3}$ was used to extract the strength of directed flow at midrapidity for $\pi^{\pm}, K^{\pm}, K^{0}_{S}, p$, and $\Lambda$, while the fit form $v_{1}(y)=by$ was used for $\phi$ and $\Xi^{-}$ due to the limited statistics. The fit range for all particles is $-$0.75 $\textless$ $y$ $\textless$ 0. Hereafter, we refer to $dv_{1}/dy|_{y=0}$ as the slope obtained from the above fit. The cubic fit term, $c$, can reduce the sensitivity to the rapidity range. The constant term, $a$, accounts for the effects from event plane fluctuation and momentum conservation~\cite{PhysRevC.66.014901}. The constant term, $a$, is found to be $<$ 0.005 for all particles except $\phi$ and $\Xi^{-}$ in the 10-40\% centrality. 

%%%%%%%%%%%%%% Fig. 3 %%%%%%%%%%%%%%%%%%%%
\begin{figure}[!htb]
\centering
\hspace*{-8mm}
\centerline{\includegraphics[width=0.53\textwidth]{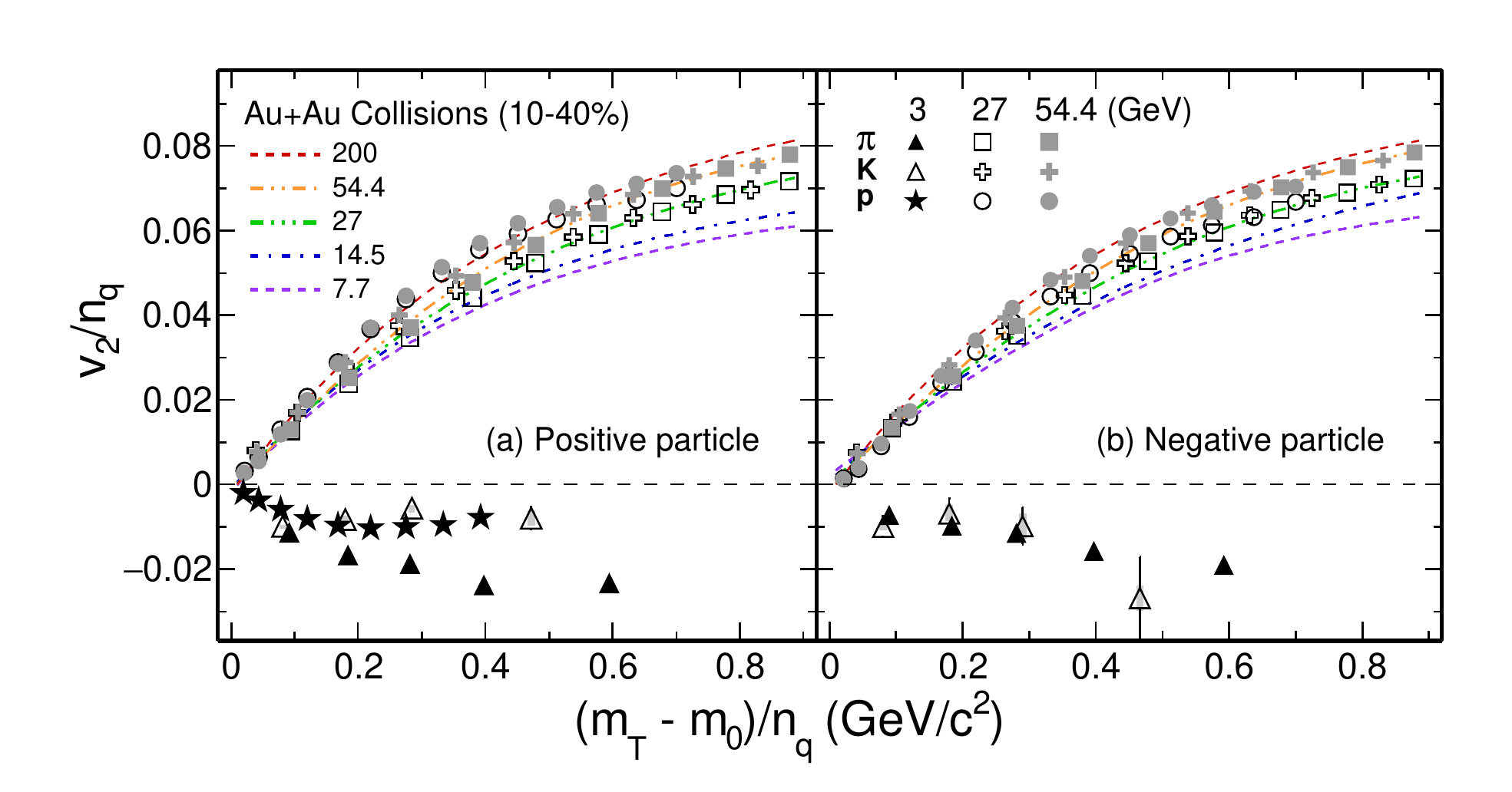}}
\caption{$v_2$ scaled by the number of constituent quarks, $v_2/n_{q}$, as a function of scaled transverse kinetic energy ($(m_{T}-m_{0})/n_{q}$) for pions, kaons and protons from Au+Au collisions in 10-40\% centrality at \snn = 3, 27, and 54.4\,GeV for positive charged particles (left panel) and negative charged particles (right panel). The measurements are in the rapidity range $|y| <$ 0.5 at 27 and 54.4 GeV, and in $-$0.5 $<y<$ 0 at 3 GeV. Colored dashed lines represent the scaling fit to data from Au+Au collisions at 7.7, 14.5, 27, 54.4, and 200\,GeV from STAR experiment at RHIC~\cite{Adamczyk:2013gv,Adamczyk:2015fum,Abelev:2008ae}. Statistical and systematic uncertainties are shown as bars and gray bands, respectively. Some uncertainties are smaller than the data points.}
\label{fig3}
\end{figure}
%%%%%%%%%%%%%%%%%%%%%%%%%%%%%%%%%%%%%

The elliptic flow scaled by the number of constituent quarks, $v_2/n_q$, for the copiously produced hadrons $\pi^{\pm}$ (squares), $K^{\pm}$ (crosses), $p$ and $\bar{p}$ (circles) is shown as a function of the scaled transverse kinetic energy $(m_T-m_0)/n_q$ in Fig.~\ref{fig3}. Data are from 10-40\% mid-central Au+Au collisions at RHIC. Data points from collisions at 27 and 54.4\,GeV are shown as open and closed symbols, respectively. The colored dashed lines, also displayed in the figure, represent the scaling fit to data for pions, kaons, and protons in Au+Au collisions at $\sqrt{s_{NN}}$ = 7.7, 14.5, 27, 54.4, and 200\,GeV~\cite{Dong:2004ve,Adamczyk:2013gw} for both positive and negative charged particles. 
Although the overall quark number scaling is evident, it has been observed that the best scaling is reached in the RHIC top energy \snn = 200\,GeV collisions~\cite{Adamczyk:2015ukd,PHENIX:2006dpn}. As the collision energy decreases, the scaling deteriorates. Particles and antiparticles are no longer consistent with the single-particle NCQ scaling~\cite{Adamczyk:2013gw} due to the mixture of the transported and produced quarks. More detailed discussions on the effects of transported quarks on collectivity can be found in Refs.~\cite{Adamczyk:2017nxg,Dunlop:2011cf}.  As one of the important evidence for the QGP formation in high energy collisions at RHIC, the observed NCQ scaling originates from partonic collectivity~\cite{Molnar:2003ff,Adamczyk:2015ukd,Adamczyk:2017xur}. 

For Au+Au collisions at 3\,GeV, data points for $\pi$, $K$ and $p$ are represented by filled triangles, open triangles and filled stars, respectively in Fig.~\ref{fig3}. It is apparent that all of the values of $v_2/n_q$ are negative. Only proton results are shown, because of the lack of antiproton production at this energy. Contrary to the higher energy data shown, the quark scaling disappears in the observed elliptic flow for positively charged particles in such low energy collisions. The new results clearly indicate different properties for the matter produced. As shown in Fig.~\ref{fig2}, the JAM and UrQMD model calculations with baryonic mean-field potential reproduce the observed negative values of $v_2$ for protons as well as $\Lambda$s. In other words, in the Au+Au collisions at 3\,GeV, partonic interactions no longer dominate and baryonic scatterings take over. 
This observation is clear evidence that predominantly hadronic matter is created in such collisions.

%%%%%%%%%%%%%% Fig. 4 %%%%%%%%%%%%%%%%%%%%
\begin{figure}[!htb]
\centering
\centerline{\includegraphics[width=0.4\textwidth]{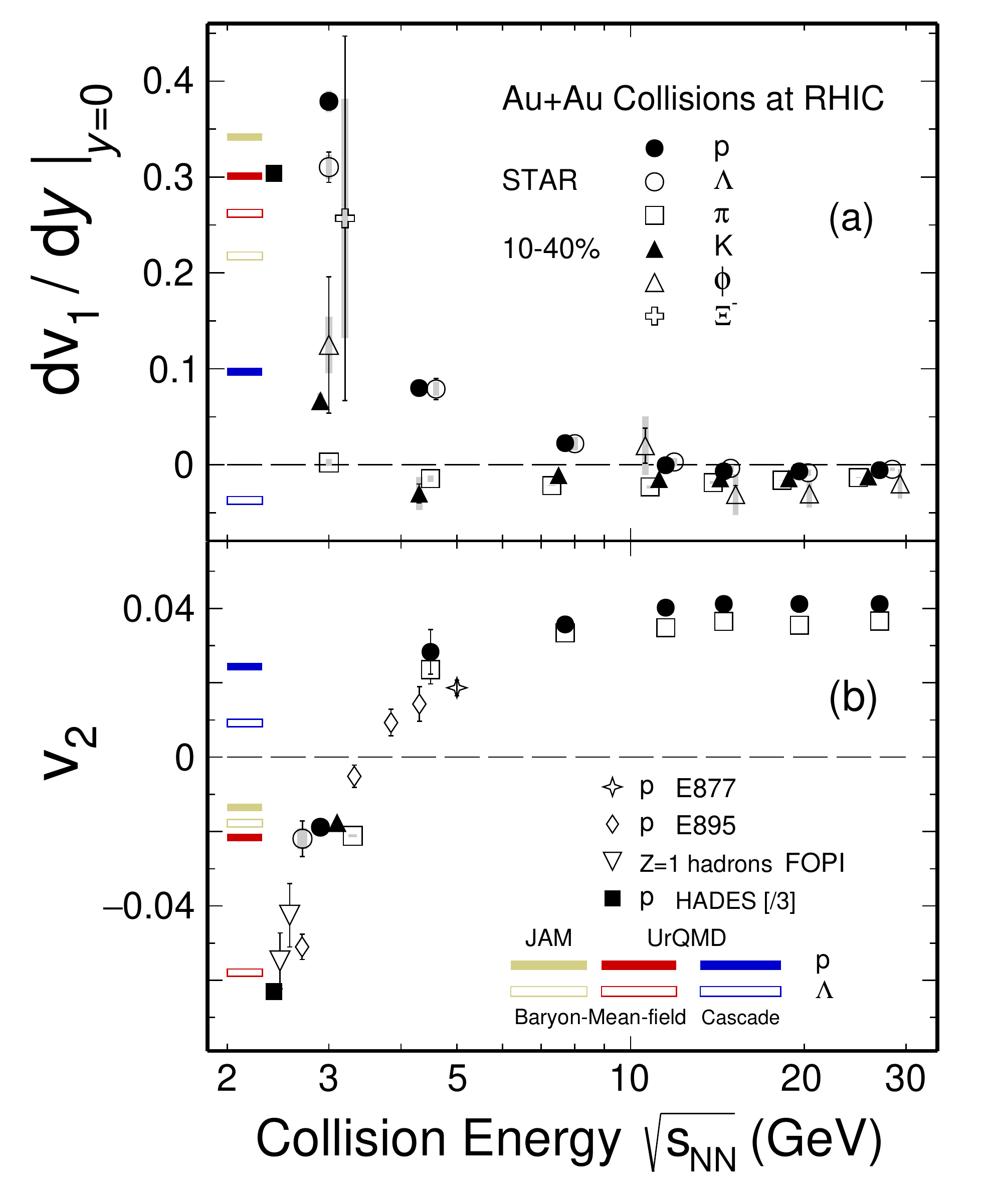}}
\caption{Collision energy dependence of (top panel) directed flow slope $dv_{1}/dy|_{y=0}$ for $p$, $\Lambda$, $\pi$(combined from $\pi^{\pm}$), $K$(combined from $K^{\pm}$ and $K^{0}_{S}$), $\phi$ and $\Xi^{-}$, and (bottom panel) elliptic flow $v_2$ for $p$, $\pi$(combined from $\pi^{\pm}$) in heavy-ion collisions~\cite{Adamczyk:2017nxg,Adamczyk:2013gw,Adam:2020pla,Pinkenburg:1999ya,Andronic:2004cp}. Collision centrality for all data from RHIC is 10-40\%, except for 4.5\,GeV where 10-30\% is for $dv_{1}/dy|_{y=0}$ and 0-30\% is for $v_{2}$. Note that the HADES~\cite{Adamczewski-Musch:2020iio} results of proton $dv_{1}/dy|_{y=0}$ and $v_2$ from 20-30\% scaled by a factor of 3 are from 1 $\textless$ $p_{T}$ $\textless$ 1.5 (GeV/$c$), which is in a higher $p_T$ region compared to our data (0.4 $\textless$ $p_{T}$ $\textless$ 2.0 (GeV/$c$)). Statistical and systematic uncertainties are shown as bars and gray bands, respectively. Some uncertainties are smaller than the data points. The JAM and UrQMD results are shown as colored bands: golden, red and blue bands stand for JAM mean-field, UrQMD mean-field and UrQMD cascade mode, respectively. 
For clarity the x-axis value of the data points have been shifted. }
\label{fig4}
\end{figure}

Collision energy dependence of the directed and elliptic flow is summarized in Fig.~\ref{fig4}, where panel (a) shows the slope of the $p_T$-integrated directed flow at midrapidity, $dv_{1}/dy|_{y = 0}$, for $\pi$, $K$, $p$, $\Lambda$ and multistrange hadrons $\phi$ and $\Xi^{-}$ from Au+Au collisions for the 10-40\% centrality interval. Here $K$ and $\pi$ are the results of combination of $K^{\pm}$ plus $K^{0}_{\rm{S}}$ and $\pi^{\pm}$, respectively. The $p_T$-integrated $v_2$ at midrapidity  of $\pi$, $K$, $p$ and $\Lambda$ are shown in panel (b) as open squares, filled triangles, filled circles and open circles, respectively. 
The recent HADES proton $v_2$ from 2.4 GeV Au+Au collisions shown by a filled square is much more negative ($\sim-$0.2) implying stronger shadowing effect at lower center of mass energy. An additional reason for the significant decrease in $v_2$ is that the $p_T$ region of HADES results is relatively higher than STAR results.
Due to partonic collectivity in Au+Au collisions at high energy~\cite{Snellings:1999bt}, all observed $v_1$ slopes and $v_2$ at midrapidity are found to be negative and positive, respectively, while the observed trend in Fig.~\ref{fig4} for Au+Au collisions at 3\,GeV is exactly the opposite. The early strong partonic expansion leads to the positive $v_2$ with NCQ scaling in high energy collisions while at 3 GeV, both weaker pressure gradient and the shadowing of the spectators result in the negative $v_2$  where the scaling is absent.
Results from calculations using the hadronic transport model JAM and UrQMD, with the same centrality and kinematic cuts as used in the data analysis, are also shown as colored bands in the figure. By including the baryonic mean-field, the JAM and UrQMD model reproduced the trends for both $dv_{1}/dy|_{y = 0}$ and $v_2$ for baryons including protons and $\Lambda$. 
The consistency of transport models (JAM and UrQMD) with baryonic mean-field for all measured baryons implies that
the dominant degrees of freedom at collision energy of  3\,GeV are the interacting baryons.
The signatures for the transition from partonic dominant to hadronic and to baryonic dominant regions have also been discussed in Ref.~\cite{Bzdak:2019pkr,Adamczyk:2014ipa,Adamczyk:2017iwn,Adamczyk:2017nxg} for the ratios of $K^+/\pi^+$ and net-particle $v_1$ slopes, respectively. Our new data clearly reveals that baryonic interactions dictate the collision dynamics in Au+Au collisions at 3\,GeV.

\section{Summary}
In summary, we have reported on the $p_{T}$ and rapidity differential and integral measurements for directed flow $v_{1}$ and elliptic flow $v_{2}$ of identified hadrons $\pi^{\pm}$, $K^{\pm}$, $K_{S}^{0}$, $\phi$, $p$, $\Lambda$ and $\Xi^{-}$ from the 10-40\% centrality Au+Au collisions at \snn = 3\,GeV, and the high statistics measurements for $v_{2}$ of $\pi^{\pm}$, $K^{\pm}$, $p$ and $\overline{p}$ at \snn = 27 and 54.4\,GeV. The NCQ scaling of $v_2$ is observed for collision energies $\ge$ 7.7\,GeV. Due to the formation of the QGP at center-of-mass collision energies larger than 10\,GeV, one finds that each hadron's $v_2$ is positive while all slopes of $v_1$ are negative. 
For Au+Au collisions at 3\,GeV, the NCQ scaling is absent and the opposite collective behavior is observed: the  elliptic flow of all hadrons at midrapidity is negative; the slope of the directed flow of all hadrons, except $\pi^+$, at midrapidity is positive. 
Furthermore, transport models JAM and UrQMD calculations with a baryonic mean-field qualitatively reproduced these results. These observations imply the vanishing of partonic collectivity and a new EOS, likely dominated by baryonic interactions in the high baryon density region.

\section*{Acknowledgement}
We thank Drs. Y. Nara and J. Steinheimer for interesting discussions and the use of their JAM and UrQMD simulations codes. We thank the RHIC Operations Group and RCF at BNL, the NERSC Center at LBNL, and the Open Science Grid consortium for providing resources and support.  This work was supported in part by the Office of Nuclear Physics within the U.S. DOE Office of Science, the U.S. National Science Foundation, the Ministry of Education and Science of the Russian Federation, National Natural Science Foundation of China, Chinese Academy of Science, the Ministry of Science and Technology of China and the Chinese Ministry of Education, the Higher Education Sprout Project by Ministry of Education at NCKU, the National Research Foundation of Korea, Czech Science Foundation and Ministry of Education, Youth and Sports of the Czech Republic, Hungarian National Research, Development and Innovation Office, New National Excellency Programme of the Hungarian Ministry of Human Capacities, Department of Atomic Energy and Department of Science and Technology of the Government of India, the National Science Centre of Poland, the Ministry  of Science, Education and Sports of the Republic of Croatia, RosAtom of Russia and German Bundesministerium f\"ur Bildung, Wissenschaft, Forschung and Technologie (BMBF), Helmholtz Association, Ministry of Education, Culture, Sports, Science, and Technology (MEXT) and Japan Society for the Promotion of Science (JSPS).

%\begin{thebibliography}
%\bibliographystyle{elsarticle-num} 
%\bibliography{reference}
%\end{thebibliography}

\end{document}